\documentclass[aps,amsmath,amssymb,showpacs,floatfix,twocolumn]{revtex4}
\usepackage{graphicx}
\usepackage{dcolumn}
\usepackage{bm}
\usepackage{hyperref} 
\usepackage{latexsym}
\usepackage{float}

\def\av#1{\langle#1\rangle}
\def\Pr{{\rm Pr}}

\begin{document}
\title{Synchronous and Asynchronous Recursive Random Scale-Free Nets}   

\author{Francesc Comellas}
\email{comellas@mat.upc.es}
\affiliation{Departament de Matem\`atica Aplicada IV, EPSC, Universitat Polit\`ecnica de Catalunya\\
Avinguda del Canal Ol\'\i mpic s/n, 08860 Castelldefels, Barcelona, Catalonia, Spain}
\author{Hernan D.~Rozenfeld}
\email{rozenfhd@clarkson.edu}
\affiliation{Department of Physics, Clarkson University,
Potsdam NY 13699-5820, USA}
\author{Daniel ben-Avraham}
\email{benavraham@clarkson.edu}
\affiliation{Department of Physics, Clarkson University,
Potsdam NY 13699-5820, USA}

\begin{abstract}
We investigate the differences between scale-free recursive nets constructed by a synchronous,
deterministic updating rule (e.g., Apollonian nets), versus an asynchronous, random sequential updating rule (e.g., random Apollonian nets).  We show that the dramatic discrepancies observed recently for the degree exponent in these two cases result from a biased choice of the units to be updated sequentially in the asynchronous version.
\end{abstract}


\maketitle

Stochastic scale-free graphs have been found in a host of natural and manmade phenomena (the internet and the World Wide Web, networks of flight connections, of social contact, of predator prey, of metabolic reactions, etc.) and have attracted much recent attention~\cite{AlBa02,DoMe02,Ne03,JeToAlOlBa00,GoOhJeKaKi02,GuMoTuAm05}.  In this context, deterministic scale-free graphs have been very useful as exactly solvable models of the stochastic scale-free nets encountered in everyday life~\cite{DoGoMe02,CoFeRa04,AnHeAnSi05,DoMa05,CoSa02}.  Most deterministic scale-free nets are recursive: they are constructed by repeated iteration of a fixed set of rules.  It is, indeed, their recursive character that makes them particularly amenable to analysis.  

Recently, studies of random variations of deterministic nets have yielded interesting results.  Consider, for example, the deterministic Apollonian net.  Its construction begins with 3 interconnected nodes (a complete graph of order 3, K$_3$) and at each iteration a new node is connected  to the vertices of {\it every\/} existing $3$-clique (a K$_3$ subset), omitting those 3-cliques that had already been updated in previous iterations.
In the random version the  3-cliques to be updated are selected at random, {\it one at a time\/}.   In both cases there results a scale-free graph, but the deterministic net has degree exponent $\gamma=1+\ln3/\ln2$~\cite{AnHeAnSi05,DoMa05,ZhYaWa05,ZhCoFeRo05}, as opposed to $\gamma=3$ of the random construct~\cite{Zh04,ZhRoCo05}.  Similar discrepancies are observed for the deterministic network of Dorogovtsev, Goltsev, and Mendes~\cite{DoGoMe02}  (where at each iteration a node is connected to the endpoints of each existing link) versus the random version (nodes are connected to the endpoints of randomly selected links, one at a time), as well as for other types of recursive graphs~\cite{other}. 
Such dramatic differences between {\it synchronous\/} (or parallel) versus {\it asynchronous\/} (or random sequential) updating are surprising.  Barring accidental symmetries, in dynamical lattice models and interacting particle systems the two kinds of updating usually yield the same kind of behavior and  critical exponents~\cite{synchronous}.

In this paper we argue that the reason behind the dramatic differences between synchronous and asynchronous updating stem from a biased choice of the units to be updated in the sequential constructions.  We focus on a particularly simple scale-free tree whose deterministic construction involves connecting new nodes to the endpoints of each of its links.  We then consider two methods of sequential updating that differ from each other only in the selection rule of the links to be updated.  The degree exponents obtained in these two ways differ from each other (as well as from that of the deterministic tree), demonstrating our point.  A third method of asynchronous updating is presented, that avoids the pitfall of biased selection and yields the same degree exponent as for synchronous updating.  We briefly explore the differences between random and deterministic constructs even when the degree exponents obtained in the two methods agree with one another.

\section*{Synchronous Updating} 
\begin{figure}[ht]
\vspace*{0.cm}
 \includegraphics*[width=0.40\textwidth]{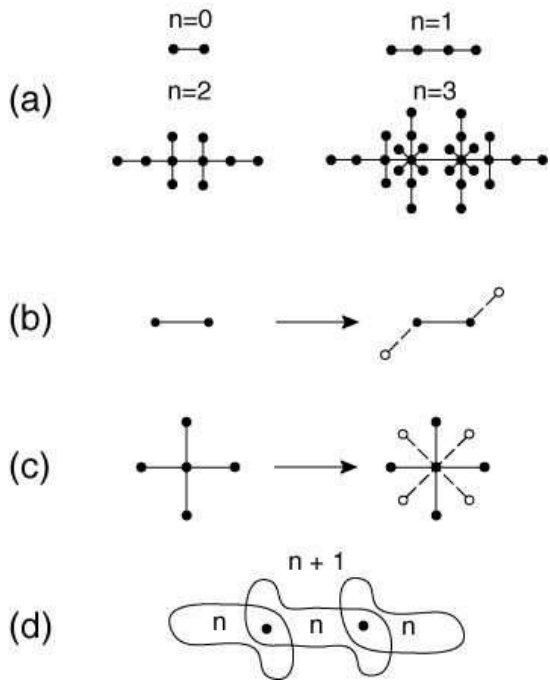}
\caption{Synchronous recursive scale-free tree and methods of construction.
(a)~Generations $n=0,1,2,3$ of the tree.
(b)~First method of construction: to each of the endpoints of every link in generation $n$ 
connect a node of degree one.
(c)~Second method: to each node of degree $k$ in generation $n$ 
add $k$ new
nodes of degree one.  
(d)~Third method: to obtain generation $n+1$, join three copies of generation $n$ at
the hubs (the nodes of highest degree).
}
\label{trees}
\end{figure}

Consider the deterministic tree of Fig.~\ref{trees}a, obtained by the following procedure~\cite{Bollt05}:  Starting from K$_2$ at generation $n=0$, construct successive generations by attaching nodes of degree one to the endpoints of each existing link (Fig.~\ref{trees}b).  The tree that emerges in generation $n$ has two hubs (the nodes of highest degree) of degree $2^n$.  An alternative way for constructing the tree consists of doubling the degree of each existing node, from $k$ to $2k$, by attaching to it $k$ single-degree nodes (Fig.~\ref{trees}c).  Yet a third method, which highlights the self-similarity  of the tree,  consists  of producing $3$ replicas of generation $n$ and joining them at the hubs (Fig.~\ref{trees}d). 

It is clear that all nodes have degrees that are powers of $2$.  Let $N_n(m)$ be the number of nodes of degree $2^m$ in generation $n$.  Let $N_n=\sum_mN_n(m)$ be the total number of nodes (the order) in generation $n$.  Let $M_n$ be the number of links (the size) in generation $n$.  We have, 
\begin{equation}
M_n=3M_{n-1}\,,\qquad M_0=1\;,
\end{equation}
(seen most easily by the first method of construction),
from which follows that
\begin{equation}
M_n=3^n\;.
\end{equation}
Since the graph is a tree,
\begin{equation}
N_n=M_n+1=3^n+1\;.
\end{equation}
Also,
\begin{equation}
N_n(m)=N_{n-1}(m-1)+2\cdot3^{n-1}\delta_{m,0}\;,
\end{equation}
leading to
\begin{equation}
\label{eq5}
N_n(m)= 
\begin{cases}
2\cdot3^{n-m-1}, & m<n\;,\\
2, & m=n\;,\\
0, & m>n\;.
\end{cases}
\end{equation}
This corresponds to a scale-free degree distribution of degree exponent $\gamma=1+\ln3/\ln 2$.

\section*{Asynchronous Updating}

Let us now explore the consequences of asynchronous, or random sequential updating, and how it differs from synchronous updating.  Based on the first method of construction (Fig.~\ref{trees}b), at each time step, $t$, we choose a link, randomly, and connect a new node to each of its endpoints.  We intend to show that the differences between synchronous and sequential updating arise because of biases in the selection rule of the link to be updated.  To this end we consider the following two rules: (a)~Select one of the $M(t)$ links in the net, randomly, with equal probability, or (b)~Select a node (among the $N(t)$ nodes of the net), randomly, then select one of its neighbors, at random, and pick the link that connects the two nodes.  A detailed analysis of these two rules and how they differ in the selection of specific links is given in Appendix~\ref{howtopick}.

Consider how the degree of node $i$, $k_i(t)$, changes with time, by method (a).  Since each of the $k_i$ links leading to node $i$ is selected with probability $1/M(t)$, the probability that $k_i\to k_1+1$ in the next time step is $k_i(t)/M(t)$.  At each time step we add two links to the tree, so $M(t)=2t-1$ (we begin with a single link at time step $t=1$).  Hence, in the long time asymptotic limit changes in $k_i$ are given by
\begin{equation}
\frac{dk_i}{dt}=\frac{k_i}{2t}\;.
\end{equation}
The initial condition for this equation is $k_i(t_i)=1$, that is, we assume that the node was introduced to the tree at time $t_i$ as a node of degree one (like all newly introduced nodes).  The solution is then
\begin{equation}
\label{ki}
k_i=\sqrt{\frac{t}{t_i}}\;.
\end{equation}
It follows from (\ref{ki}) that the probability that $k_i$ is larger than $k$, is
\begin{equation}
\chi(k)\equiv\Pr(k_i>k)=\Pr\left(t_i<\frac{t}{k^2}\right)\;.
\end{equation}
However, since node $i$ could be introduced in any of the $t$ steps with equal probability, the probability that $t_i<T$ is $T/t$, so that $\chi(k)=1/k^2$.
The degree distribution for large $k$ then follows:
\begin{equation}
P_a(k)=-\frac{d}{dk}\chi(k)\sim k^{-3}\;.
\end{equation}
We conclude that the random sequential construction of the tree by method~(a) leads to a scale-free degree distribution of degree exponent $\gamma=3$, different from $\gamma=1+\ln3/\ln2$ of the deterministic tree.

But what if we select the links by method~(b)?  In this case node $i$ may be the first of the two nodes to be selected in step $t$.  This may happen with probability $1/N(t)$.  Node $i$ may also be the second node selected.  The probability that we reach $i$ through a randomly selected node is $k_i/N\av{k}$.  The degree of node $i$ increases from $k_i$ to $k_{i}+1$ regardless of whether it is picked first or second.  That is, the rate of increase is $(1/N)(1+k_i/\av{k})$.  But $N(t)=2t$, while $\av{k}=2M/N=(4t-2)/(2t)\to2$, as $t\to\infty$.
It follows that in the long time asymptotic limit
\begin{equation}
\frac{dk_i}{dt}=\frac{1}{2t}\left(1+\frac{k_i}{2}\right)\;.
\end{equation}
From here we proceed exactly as for method~(a), this time obtaining
\begin{equation}
P_b(k)\sim k^{-5}\;,
\end{equation}
for large $k$.  Once again the degree distribution is scale-free.  The degree exponent $\gamma=5$ is not only different from that of the deterministic tree but also differs from $\gamma=3$ obtained by method~(a).

\begin{figure}[ht]
 \vspace*{0.cm}\includegraphics*[width=0.40\textwidth]{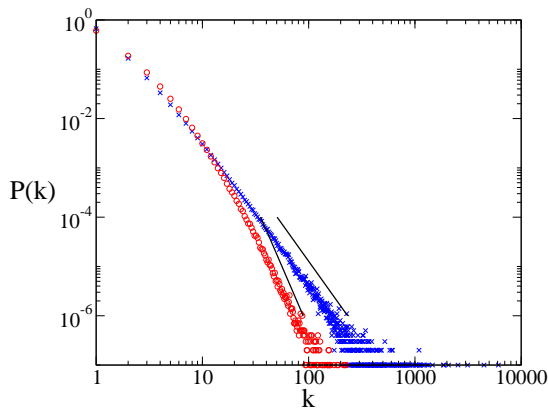}
\caption{
Random sequential trees constructed by 
selecting links at random ($\times$), or by picking the link between a randomly selected node 
and a neighbor ($\circ$).  The trees consist of $10^7$ nodes.  
The straight lines of slope 3 and 5 are shown for comparison.
}
\label{Pk}
\end{figure}

In Fig.~\ref{Pk} we show the degree distribution of random sequential trees constructed by the two methods.
The simulation data are consistent with the degree exponents predicted above, though one could argue that larger simulations are needed for the distributions to converge to their long time asymptotic limit~\cite{remark1}.  At any rate, the simulations demonstrate our point that a bias in the selection of the updated units is responsible for the differences between different kinds of random sequential nets,  and 
between random sequential nets and synchronous nets.  The differences from the deterministic tree, in our case,
result from the fact that picking links at random (by either of the two methods considered here) favors links with endpoints of higher degree.

A third method for building the random sequential tree is based on the recursive technique of Fig.~\ref{trees}c, where at each iteration the degree of all existing nodes is doubled by attaching to them new nodes of degree one.  In the random sequential version the degree of only one randomly selected node is doubled at each time step.  The node to be updated is picked with equal probability from among the $N(t)$ nodes of the net.  We now show that a random net constructed by this method achieves the same degree exponent as the deterministic net.

We cannot resort to the same technique employed for the previous two random sequential constructs because 
now $k_i\to2k_i$ in a single step, and the increase is too large to allow for a continuous  approximation (we are trying to study $k_i$ large).  Instead, we resort to discrete rate equations.  

The degrees in the random tree are powers of 2, just as in the deterministic tree.   Let $N_t(m)$ be the number of nodes of degree $2^m$ at time $t$, and let $N_t=\sum_mN_t(m)$ be the
order of the tree, then
\begin{eqnarray}
\label{eq6}
\begin{aligned}
&N_{t+1}(m)= N_t(m)+\frac{N_t(m-1)}{N_t}-\frac{N_t(m)}{N_t}\;,\quad m\geq1,\\
&N_{t+1}(0)= N_t(0)-\frac{N_t(0)}{N_t}+\sum_{m=0}2^m\frac{N_t(m)}{N_t}\;.
\end{aligned}
\end{eqnarray}
The long time asymptotic limit can be obtained by making the ansatz: $N_t(m)\to b_mt$
and $N_t\to Bt$, as $t\to\infty$ ($B=\sum_mb_m$).
Substituting in the first of Eqs.~(\ref{eq6}), we learn that $b_m=b_{m-1}/(1+B)$, leading to
$b_m=b_0/(1+B)^m$.  The second equation tells us that $B=2$, and the remaining $b_0$ is obtained from the constraint $\sum_mb_m=B$. We get
\begin{equation}
N_t(m)\to \frac{4}{3^{m+1}}\,t\;.
\end{equation}
This is essentially the same distribution as~(\ref{eq5}), corresponding to $\gamma=1+\ln3/\ln2$.  This shows that there exist ways of choosing the units to be updated such that the differences between random sequential nets and deterministic nets are minimal.

\begin{figure}[ht]
 \vspace*{0.cm}\includegraphics*[width=0.40\textwidth]{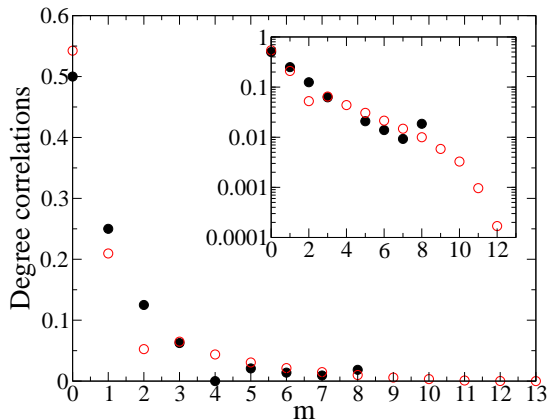}
\caption{
Degree-degree correlations in deterministic and random trees.   
Shown is $M(4,m)/\sum_kM(4,k)$ as a function of $m$ for deterministic trees of generation $n=8$ ($\bullet$), compared to random trees of equal size ($\circ$).  The random results were obtained as an average over 1000 realizations.  Error bars are smaller than the size of the symbols.
}
\label{correlations}
\end{figure}

An interesting question concerns the differences between the deterministic and random scale-free nets, even when the random construction is unbiased.  
It is evident, from our last example, that differences do exist.  For instance, in the deterministic tree there are exactly two hubs (the highest degree nodes), always connected to one another.  This need not be the case in the corresponding random tree.  More generally, let $M_n(k,l)$ be the number
of links with endpoints of degree $2^k$, $2^l$ for generation $n$ of the deterministic tree.  Then
\begin{equation}
\begin{aligned}
&M_n(k,l)=M_{n-1}(k-1,l-1)\;, \qquad 0<k\leq l\;,\\
&M_n(0,l)=2^{l-1}N_{n-1}(l-1)\;,
\end{aligned}
\end{equation}
leading to
\begin{equation}
M_n(k,l)=\begin{cases}
2^{l-k}3^{n-l-1}, & k<l<n,\\
2^{n-k}, & k\leq l=n.
\end{cases}
\end{equation}
This can be compared to $M_t(k,l)$ of random trees of an equivalent size (Fig.~\ref{correlations}).
The differences seem to be trivial, leaving us with the question whether significant discrepancies show with respect to any other measure of structure.

\acknowledgments

DbA gratefully acknowledges partial support from NSF award PHY0140094.
Support for F.C. was provided by the Secretaria de Estado de Universidades
e Investigaci\'on (Ministerio de Educaci\'on y Ciencia),  Spain, and the
European Regional Development Fund (ERDF) under project TIC2002-00155.

\appendix
\section{On Picking a Random Link}
\label{howtopick}
How does one pick a link at random?  We focus on the following two methods:  (a)~Select a link at random from among all the links in the network, with equal probabilities, or (b)~Select a node at random, then select a neighbor of that node at random, and pick the link connecting the two nodes.  The two methods select specific links with generically different probabilities. 

Consider a particular link, in a graph of order $N$ and size $M$, whose endpoints are nodes of degree $k_1$ and $k_2$.  In method (a) each of the $M$ links in the graph is selected with equal probability, so the link under consideration is selected with probability 
\begin{equation}
p_{\rm link}=\frac{1}{M}\;.
\end{equation}
In method (b), the probability to pick the node of degree $k_1$ first, is $1/N$.  Then the probability that the right neighbor is selected, is $1/k_1$.  Likewise, the probability to select the node of degree $k_2$ first is $1/N$, and the remaining node is selected with probability $1/k_2$.  It follows that the probability to hit upon the specified link by random selection of the nodes at its endpoints is
\begin{equation}
p_{\rm nodes}=\frac{1}{N}\left(\frac{1}{k_1}+\frac{1}{k_2}\right)\;.
\end{equation}
Since $2M/N=\av{k}$, the average degree of nodes in the graph, it follows that
\begin{equation}
\label{p/p}
\frac{p_{\rm link}}{p_{\rm nodes}}=\frac{\av{k}}{2}\left(\frac{1}{k_1}+\frac{1}{k_2}\right)\;.
\end{equation}
Note that this relation is valid for all connected graphs, regardless of structural details.
An example relevant to our paper is a tree graph, where $M=N-1$, and $\av{k}\approx2$ if the tree is large.
In this case we deduce from~(\ref{p/p}) that only links with $k_1=k_2=2$ are equally likely to be picked by either method.  Method~(b) favors links with $\min\{k_1,k_2\}>2$.  On the other hand, all the tree leaves (links leading to a node of degree one) are selected more
frequently by method~(a).


\begin{thebibliography}{99}
\bibitem{remark1} Indeed, smaller simulations of nets with a few thousand nodes fail to reveal convincing differences between the two methods and the degree exponent seems then consistent with that of the deterministic tree.  Note that even with $10^7$ nodes the largest degree in the nets are only about 100 (method b) and 300 (a).
\end{thebibliography}
\end{document}